\documentclass[aps,showpacs,superscriptaddress,tightenlines,a4paper]{revtex4}

\usepackage{amssymb}
\usepackage{amsmath}
\usepackage{bbold}
\usepackage{mathrsfs}
\usepackage{paralist}
\usepackage{graphicx}
\usepackage{xcolor}

\newcommand{\ii}{{\rm i}}
\newcommand{\e}{{\rm e}}

\begin{document}
%
%
    \title{Negative frequencies in wave propagation: a microscopic model}
%
%
	\author{S. A. R. Horsley}
	\affiliation{Department of Physics and Astronomy, University of Exeter,
Stocker Road, Exeter, EX4 4QL}
	\email{s.horsley@exeter.ac.uk}
	
	\author{S. Bugler--Lamb}
	\affiliation{Department of Physics and Astronomy, University of Exeter,
Stocker Road, Exeter, EX4 4QL}
	\email{slb235@exeter.ac.uk}
%
%
    \begin{abstract}
    A change in the sign of the frequency of a wave between two inertial reference frames corresponds to a reversal of the phase velocity.  Yet from the point of view of the relation \(E=\hbar\omega\), a positive quantum of energy apparently becomes a negative energy one.  This is physically distinct from a change in the sign of the wave--vector, and has been associated with various effects such as Cherenkov radiation, quantum friction, and the Hawking effect.  In this work we provide a more detailed understanding of these negative frequency modes based on a simple microscopic model of a dielectric medium as a lattice of scatterers.  We calculate the classical and quantum mechanical radiation damping of an oscillator moving through such a lattice and find that the modes where the frequency has changed sign contribute negatively.  In terms of the lattice of scatterers we find that this negative radiation damping arises due to phase of the periodic force experienced by the oscillator due to the relative motion of the lattice.
    \end{abstract}
%
%
    \pacs{03.50.De,81.05.Xj, 78.67.Pt}
    \maketitle
%
%
	\par
	\section{\label{sec:Introduction}Introduction}
	\par
	Despite the lack of any immediate physical interpretation, changing the sign of the frequency \(\omega\to-\omega\) of a classical wave from a positive to a negative number does not contradict any physical law.  For instance, a plane wave \(\varphi\) moving with a phase velocity \(v_{p}=\omega/k\) can be written as,
	\begin{equation}
		\varphi(x,t)=\cos(k x\pm\omega(k) t)=\cos[k( x\pm v_{p} t)]\label{classical-wave}
	\end{equation}
	where the frequency \(\omega(k)\) is a function of wavevector \(k\).  All that happens when \(\omega\to-\omega\) is that the wave reverses its direction, \(v_{p}\to-v_{p}\).  In this respect the relative sign of \(\omega\) and \(k\) is all that matters, as it encodes the propagation direction of the wave.  Typically the convention is that \(\omega\) is positive so that the direction is determined by the sign of \(k\).  Yet one cannot ensure that this convention always holds, for there are often situations---usually involving a change of reference frame---where waves appear to reverse their direction through a change in sign of the frequency rather than wave--vector.  Such situations are not trivial, and they are associated with some interesting physical effects:  Cherenkov radiation is emitted in a cone consisting of waves with positive frequency in the laboratory frame, but negative frequency in the rest frame of the particle~\cite{volume8,luo2003}; the ``quantum friction'' that has been a subject of recent debate and is predicted to occur between relatively moving dielectrics at zero temperature~\cite{pendry1997,volokitin2007} has similar origin to the Cherenkov effect~\cite{volume8,maghrebi2013,maslovski2013}, and stems from the mixing of positive and negative frequency waves in the two materials~\cite{horsley2012,guo2014}; and Hawking radiation~\cite{hawking1974} and its laboratory analogues~\cite{barcelo2005,philbin2008} originate from the change in the sign of the frequency of a wave as it crosses the event horizon, an effect that has been recently experimentally observed in a number of analogue systems~\cite{rousseaux2008,belgiorno2010,weinfurtner2011,steinhauer2014}, and led to the identification of new pulse propagation phenomena within fibre optics~\cite{rubino2012}.  In this paper we use a very simple model to clarify the physical meaning of a frequency that changes sign between reference frames.  We do this through considering the effect that such modes have on a moving oscillator.  The internal degree of freedom of the oscillator is imagined as a probe of the local density of states, and as a model for any system coupled to a moving medium, e.g. an atom above a moving surface~\cite{horsley2012}.
	\par
	Before introducing the model we consider an example to illustrate when we should expect frequencies to change sign between reference frames.  Imagine two cases where an observer travels relative to an electromagnetic wave: (i) moving through free space at velocity \(\boldsymbol{V}=V\hat{\boldsymbol{x}}\) relative to a light wave with a wavevector at angle \(\theta\), \(\boldsymbol{k}=(\omega/c)[\cos(\theta)\hat{\boldsymbol{x}}+\sin(\theta)\hat{\boldsymbol{y}}]\); and (ii) moving through a medium at the same velocity but parallel to a light wave \(\boldsymbol{k}=({\rm n}\omega/c)\hat{\boldsymbol{x}}\), where \({\rm n}>1\) is the refractive index of the medium.  In both cases the moving observer sees a wave with wave--vector and frequency that have undergone a Lorentz transformation~\cite{volume2}
	\begin{align}
		k_{x}'&=\gamma\left(k_{x}-\frac{\omega V}{c^{2}}\right)\nonumber\\
		\omega'&=\gamma\left(\omega-Vk_{x}\right)
	\end{align}
	where \(\gamma=(1-V^{2}/c^{2})^{-1/2}\).   In both cases (i) and (ii) the observer can travel fast enough for the wave to appear to reverse its direction of propagation.  In case (i) this occurs when \(V>c\cos(\theta)\), which is where \(k_{x}'\) takes a different sign in the frame of the moving observer, with \(\omega'\) remaining positive.  Yet in case (ii) the wave appears to reverse direction when \(V>c/{\rm n}\) where the transformed wavevector \(k_{x}'\) remains positive, but the frequency \(\omega'\) changes sign.  Consequently, even though we conventionally take all frequencies positive in one frame of reference, changing reference frame can cause some of these frequencies to change sign.  Applying the formula \(E=\hbar\omega'\), one is led to the conclusion that positive energy quanta in the rest frame appear as negative energy quanta in the moving frame.  Such negative energy quanta imply a kind of instability where waves can be excited by a moving particle, while lowering the energy of the system.  Such waves were recognised during the early work on quantum electrodynamics by authors such as Jauch and Watson~\cite{jauch1948a,jauch1948b}, but still remain somewhat mysterious.  A recent more detailed analysis that includes the dispersion and dissipation of the medium~\cite{horsley2012} shows that these modes cause the Hamiltonian of the field to lack a lower bound, and are the origin of the force of quantum friction~\cite{simpson2015}.   From this example it is clear that the change in the sign of the frequency has its origin in the inclusion of a macroscopic description of a dielectric medium through the refractive index \({\rm n}\) (the effect disappearing when \({\rm n}=1\)).  In this work we use a simple microscopic model of a dielectric medium---similar to that described in~\cite{griffiths1992,griffiths2001}---to understand the microscopic meaning of frequencies that change their sign between reference frames, and the origin of the negative energy quanta that arise when \(V>c/{\rm n}\).
%
%
%
%
\section{\label{sec:Microscopic-Model}1D model of a dielectric medium}
\par
\begin{figure}[h!]
	\includegraphics[height=5.5cm]{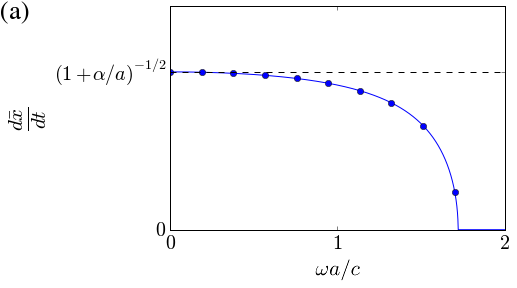}\includegraphics[height=5.5cm]{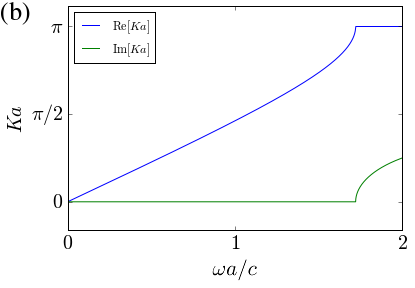}
	\caption{The velocity of energy flow and the Bloch vector \(K\) in the lattice as a function of \(\omega\), plotted for \(\alpha/a=1\).  Panel (a) shows the velocity of energy flow through the lattice (\ref{energy_velocity}) in units of \(c\) as a function of \(\omega\), for a frequency independent \(\alpha\).  In the long wavelength limit the energy propagates at the phase velocity \(\omega/k\) with the refractive index given by (\ref{macK}).  As one approaches the Brillouin zone boundary (\(K=\pi/a\)) the energy flow velocity reduces to zero.  The blue circles are the values of \(\partial\omega/\partial K\) computed from (\ref{eq:BlochVector}), showing that the velocity of energy flow equals the group velocity.  Panel (b) shows the real and imaginary parts of the Bloch wavevector as a function of \(\omega\), which is real for \(K<\pi/a\).\label{power-flow-fig}}
\end{figure}

In the spirit of taking simplest cases first, we consider a one--dimensional model for a dielectric medium.  For a fixed frequency \(\omega\) the waves obey the Helmholtz equation
\begin{equation}
	\left\{\frac{d^{2}}{d x^{2}}+\frac{\omega^{2}}{c^{2}}\left[1+\sum_{n}\alpha\delta(x-na)\right]\right\}\varphi(x,\omega)=0\label{eq:WaveEq}
\end{equation}
where the medium is taken to be composed of equally spaced point--like scatterers (atoms) at positions \(x=na\), each with a polarizability \(\alpha\) (in this work \(\alpha\) is independent of frequency).  At each atom position the wave is continuous
\begin{equation}
	\lim_{\eta\to0}\left[\varphi(na-\eta)-\varphi(na+\eta)\right]=0\label{bc1}
\end{equation}
Meanwhile the boundary condition on the derivative of \(\varphi\) can be found through integrating (\ref{eq:WaveEq}) over the infinitesimal interval \([na-\eta,na+\eta]\)
\begin{equation}
	\lim_{\eta\to0}\left[\frac{\partial\varphi}{\partial x}\right]_{na-\eta}^{na+\eta}\\
	=-\frac{\omega^{2}}{c^{2}}\alpha\varphi(na)\label{bc2}.
\end{equation}
Either side of \(x=na\) the field is a sum of two plane waves with wavevector \(\pm \omega/c\).  These waves are subject to the boundary conditions (\ref{bc1}--\ref{bc2}) along with the periodic boundary condition \(\varphi(x+a)=\exp(\ii K a)\varphi(x)\).  Concentrating on the unit cell centred around \(x=0\), a suitable form of the field can be written down immediately
\begin{equation}
	\varphi_{n,K}(x)=N_{n,K}\begin{cases}
		\e^{-\ii K a/2}\left[\sin((k_{n,K}+K)a/2)\e^{\ii k_{n,K}(x+a/2)}+\sin((k_{n,K}-K)a/2)\e^{-\ii k_{n,K}(x+a/2)}\right]&x<0\\[15pt]
		\e^{\ii K a/2}\left[\sin((k_{n,K}+K)a/2)\e^{\ii k_{n,K}(x-a/2)}+\sin((k_{n,K}-K)a/2)\e^{-\ii k_{n,K}(x-a/2)}\right]&x>0\label{field-form}
	\end{cases}
\end{equation}
For a fixed value of \(K\) there are many possible frequencies \(k_{n,K}=\omega/c\) that obey the dispersion relation \(\omega(K)\) and the subscript \(n\) indicates which of these we are considering.  The quantity \(N_{n,K}\) is a normalization constant that we shall determine below.  The form of the field given by (\ref{field-form}) obeys the periodic boundary condition as well as (\ref{bc1}), but we have not yet imposed (\ref{bc2}).  Imposing condition (\ref{bc2}) yields the dispersion relation that relates \(\omega\) and \(K\)
\begin{equation}
	K=\pm\frac{1}{a}\arccos\left(\cos\left(k_{n,K}a\right)-\frac{\alpha k_{n,K}}{2}\sin\left(k_{n,K}a\right)\right).\label{eq:BlochVector}
\end{equation}
The quantity \(K\) is the Bloch vector, the real part of which is proportional to the phase shift between neighbouring unit cells.  Figure~\ref{power-flow-fig} shows this quantity plotted as a function of frequency for the case of \(\alpha/a=1\).  For small \(\omega a/c\) it takes the approximate form
\begin{equation}
	K\sim\pm\sqrt{1+\frac{\alpha}{a}}\,\frac{\omega}{c}=\pm {\rm n} \frac{\omega}{c}\label{macK}
\end{equation}
which is the result that would be obtained for a macroscopic medium with the uniform permittivity \(\epsilon=1+\alpha/a\), and with index \({\rm n}=\sqrt{\epsilon}\)  (and is also the spatial average of the microscopic permittivity given in (\ref{eq:WaveEq})).  Throughout this paper we refer to this limit as the \emph{macroscopic limit}.
\par
To determine the normalization constant \(N_{n,K}\) we write the field (\ref{field-form}) in the following form
\[
	\varphi_{n,K}(x)=N_{n,K} u_{n,K}(x)\e^{\ii K x}
\]
where \(u_{n,K}\) is a periodic function of \(x\) with period \(a\).  The normalization is chosen such that
\begin{equation}
	N_{n,K}N^{\star}_{m,K}\int_{-a/2}^{a/2}\rho(x)u_{n,K}(x)u^{\star}_{m,K}(x)dx=a\delta_{nm}\label{norm-cond}.
\end{equation}
After applying the dispersion relation (\ref{eq:BlochVector}) we find this to be
\begin{equation}
	N_{n,K}=\sin^{-1/2}(k_{n,K}a)\left[\left(1+\frac{\alpha}{2a}\right)\sin(k_{n,K}a)+\frac{\alpha k_{n,K}}{2}\cos(k_{n,K}a)\right]^{-1/2}.
\end{equation}
Note that with the normalization (\ref{norm-cond}) we can construct a delta function as follows
\begin{equation}
	\int_{-\pi/a}^{\pi/a}\frac{dK}{2\pi}{\rm e}^{{\rm i}K(x-x')}\sum_{n}\rho(x')u_{n,K}(x)u_{n,K}^{\star}(x')=\delta(x-x')
\end{equation}
which is important for the construction of the quantum mechanical field operators in section~\ref{quantum-section}.
\par
To complete our description of our model medium we calculate the velocity of energy flow through the lattice.  We begin with the expression for the time--averaged power flow through the medium \(\langle S\rangle\).  For a monochromatic wave of frequency \(\omega\) this is given by
\[
	\langle S\rangle=-\left\langle\frac{\partial\varphi(x,t)}{\partial t}\frac{\partial\varphi(x,t)}{\partial x}\right\rangle=\frac{\omega}{2}{\rm Im}\left[\varphi^{\star}(x,\omega)\frac{d\varphi(x,\omega)}{dx}\right]
\]
where the units of \(\varphi\) are such that the right hand side has the dimensions of energy per unit time.  The power is independent of \(x\) for real \(\alpha\), as can be established through taking the derivative with respect to \(x\) and applying (\ref{eq:WaveEq}).  For the mode of the lattice given by (\ref{field-form}) and normalized according to (\ref{norm-cond}) the power flow is equal to
\begin{equation}
	\langle S\rangle=\frac{1}{2a}\left(\frac{\omega}{c}\right)^{2}v_{g}\label{S-lat}
\end{equation}
which is reduced relative to the power flow through free space (of a similarly normalized mode) due to the reduced group velocity \(v_{g}=d\omega/dK<c\).  As one would expect, the scatterers impede the propagation of power through the lattice.  The velocity at which this energy flows through the lattice can also be calculated in this model.  The power flow obeys a continuity equation,
\begin{equation}
	\frac{\partial\langle S\rangle}{\partial x}+\frac{\partial\langle U\rangle}{\partial t}=0\label{continuity}
\end{equation}
where the time averaged energy density \(\langle U\rangle\) is given by
\begin{equation}
	\langle U\rangle=\frac{1}{2}\left\langle\left(\frac{\partial\varphi}{\partial x}\right)^{2}+\frac{1}{c^{2}}\left[1+\alpha\sum_{n}\delta(x-n a)\right]\left(\frac{\partial\varphi}{\partial t}\right)^{2}\right\rangle\label{energy-density}
\end{equation}
which in the limit of a monochromatic field becomes
\begin{equation}
	\langle U\rangle\to\frac{1}{4}\left\{\frac{\partial\varphi}{\partial x}\frac{\partial\varphi^{\star}}{\partial x}+\frac{\omega^{2}}{c^{2}}\left[1+\alpha\sum_{n}\delta(x-n a)\right]\varphi\varphi^{\star}\right\}\label{energy-density-mono}
\end{equation}
For a frequency dependent \(\alpha\) the limiting form of \(\langle U\rangle\) given in (\ref{energy-density-mono}) would be given by the Brillouin energy density~\cite{volume8}.  Using the continuity equation (\ref{continuity}) we find that the centre of energy of the field,
\begin{equation}
	\bar{x}(t)=\frac{\int dx x \langle U\rangle}{\int dx \langle U\rangle}\label{coe}
\end{equation}
moves with a velocity given by the ratio of the total power to the total energy
\begin{equation}
	\frac{d\bar{x}(t)}{dt}=\frac{\int dx \langle S\rangle}{\int dx \langle U\rangle}.\label{coev}
\end{equation}
which can be established after an integration by parts.  Using the monochromatic modes of the lattice (\ref{field-form}) and expression (\ref{energy-density-mono}) one obtains
\begin{equation}
	\frac{d\bar{x}}{d t}=\frac{c\sin(k_{n,K}a)\sin(Ka)}{1-\cos(k_{n,K}a)\cos(K a)+\frac{\alpha}{2a}\sin^{2}(k_{n,K}a)}=v_{g}\label{energy_velocity}
\end{equation}
which reduces to \(c\) in the limit \(\alpha\to0\), and \(0\) in the limit \(\alpha\to\infty\).  The equality of the energy flow velocity with the group velocity can be found after an application of the dispersion relation (\ref{eq:BlochVector}).  Notice that the final term in the denominator comes from integrating over the delta functions in the energy density (\ref{energy-density}) and reduces the velocity at which energy can move through the lattice.  This term represents a contribution to the energy from the polarizability of the lattice itself, in addition to the effect of the scattering.  Figure~\ref{power-flow-fig}a shows the velocity (\ref{energy_velocity}) plotted as a function of frequency.
\par
As established in the introduction, the appearance of negative frequencies is ordinarily associated with a refractive index, a quantity which emerges in the macroscopic limit, which is where \(\omega a/c\ll1\) and concerns the average properties of the wave.  The spatial average of the wave over a unit cell centered on \(x\) is given by
\begin{align}
	\langle\varphi(x,t)\rangle&=\frac{{\rm e}^{-{\rm i}c k_{n,K}t}}{a}\int_{-a/2}^{a/2}\varphi(x+y)\ dy\nonumber\\
	&=\frac{{\rm e}^{{\rm i}(K x-ck_{n,K}t)}}{a}\int_{-a/2}^{a/2}u_{n,K}(x+y){\rm e}^{{\rm i}K y}\nonumber\\
	&=\e^{\ii (K x-c k_{n,K}t)}\frac{1}{a}\sum_{m=0}^{\infty}\frac{(\ii K)^{m}}{m!}\int_{-a/2}^{a/2}y^{m}u_{n,K}(x+y)\ dy\nonumber\\
	&\sim \e^{\ii (K x-ck_{n,K}t)}\langle u_{n,K}(x)\rangle=A\, \e^{\ii k_{n,K}({\rm n} x-ct)}\label{eq:AverageIntegral}
\end{align}
where \(A\) is a constant.
Thus, for small \(\omega a/c\) the average behaviour of the wave is as a plane wave with a wavevector \({\rm n}\omega/c\) where \({\rm n}\) is given by (\ref{macK}).  It is clear from figure~\ref{power-flow-fig} that in this regime---with \(\alpha\) independent of frequency---the medium behaves in a way that is equivalent to a dispersionless homogeneous medium with index \(\sqrt{1+\alpha/a}\).
%
%
\section{Wave propagation through a moving lattice, and the meaning of negative frequencies\label{neg-freq-sec}}

Having outlined the model system we now examine the change in behaviour of the waves when the lattice is set into motion, and how the coupling of wave energy from a source into the lattice changes when the velocity of the lattice is increased (in particular when \(V>c/{\rm n}\)).  In some respects this calculation is similar to those that deal with Cherenkov radiation in periodic media (e.g.~\cite{luo2003}), but differs in that we are using the periodic lattice as a model to understand the underlying physics of a \emph{uniform} medium, which emerges from the lattice model in the long wavelength limit.
\par
Performing a Lorentz transformation to a frame where the lattice is in motion with velocity \(-V\)
\begin{align}
	x&=\gamma(x'-Vt')\nonumber\\
	t&=\gamma(t'-\frac{V}{c^{2}}x').\label{lt}
\end{align}
In the unit cell centred around the origin the expression for the scalar field \(\varphi\) is given by (\ref{field-form}), and when subject to the Lorentz transformation (\ref{lt}) becomes
\begin{equation}
	\varphi(x',t')=N_{n,K}
	\begin{cases}
		{\rm e}^{-{\rm i}Ka/2}\big[\sin((k_{n,K}+K)a/2){\rm e}^{{\rm i}k_{n,K}a/2}{\rm e}^{{\rm i}k_{n,K+}(x'-ct')}\\[10pt]
		\hfill+\sin((k_{n,K}-K)a/2){\rm e}^{-{\rm i}k_{n,K}a/2}{\rm e}^{{-\rm i}k_{n,K-}(x'+ct')}\big]&-\frac{a}{2\gamma}<x'+Vt'<0\\[20pt]
		{\rm e}^{{\rm i}Ka/2}\big[\sin((k_{n,K}+K)a/2){\rm e}^{-{\rm i}k_{n,K}a/2}{\rm e}^{{\rm i}k_{n,K+}(x'-ct')}\\[10pt]
		\hfill+\sin((k_{n,K}-K)a/2){\rm e}^{{\rm i}k_{n,K}a/2}{\rm e}^{{-\rm i}k_{n,K-}(x'+ct')}\big]&0<x'+Vt'<\frac{a}{2\gamma}
	\end{cases}
\label{moving-lattice}
\end{equation}
where
\begin{align}
	k_{n,K+}&=\sqrt{\frac{1+V/c}{1-V/c}}k_{n,K}\nonumber\\
	k_{n,K-}&=\sqrt{\frac{1-V/c}{1+V/c}}k_{n,K}
\end{align}
which are the two Doppler shifted frequencies of the wave divided by \(c\).  The result (\ref{moving-lattice}) shows that in the frame where the lattice is in motion the spacing between the scatterers is contracted to \(a/\gamma\), and between each of the scatterers we have travelling waves of \emph{positive} frequencies \(ck_{n,K\pm}\).  Yet the time dependence of the field (\ref{moving-lattice}) observed at a fixed point \(x'\) has two contributions: the first is due to the factors of \(\exp(-{\rm i}c k_{n,K\pm} t)\) coming from the propagation of the waves through the empty space between the scatterers; and the second is due to the fact that the lattice is in motion, causing periodic jumps in the wave amplitude as each scatterer moves past the observer (evident in the piecewise definition of (\ref{moving-lattice})).  \emph{The negative frequencies outlined in the introduction that occur in the macroscopic limit when \(V>c/{\rm n}\) are an expression of these periodic jumps in the field}.  To see this explicitly consider the transformation of the field given expanded as a Fourier sum: \(u_{n,K}(x)=\sum_{m}u_{n,K}(m)\exp(2\pi{\rm i}m x/a)\)
\begin{align}
	\varphi(x,t)&=\sum_{m}u_{n,K}(m){\rm e}^{{\rm i}[(K+\frac{2\pi m}{a})x-\omega_{n,K} t]}\nonumber\\
	\to\varphi(x',t')&=\sum_{m}u_{n,k}(m){\rm e}^{{\rm i}\gamma[K+\frac{2\pi n}{a}-\frac{Vk_{n,K}}{c}]x'}{\rm e}^{-{\rm i}c\gamma[k_{n,K}-\frac{V}{c}(K+\frac{2\pi n}{a})]t'}\label{lt-diffracted}
\end{align}
which is a superposition of waves with frequencies \(\omega'_{n,K,m}=c\gamma[k_{n,K}-\frac{V}{c}(K+\frac{2\pi m}{a})]\), some of which have changed sign between the two reference frames.  This is true irrespective of the smallness of \(V\) because \(m\) can take arbitrarily large positive or negative values.  The terms in the sum are a Fourier representation of the piecewise definition of the field given in (\ref{moving-lattice}), and therefore---as anticipated---the negative frequencies that appear in the macroscopic limit are part of the description of the motion of the scatterers past the observer at fixed \(x'\).  When \(\omega a/c\ll 1\) then the sum (\ref{lt-diffracted}) is dominated by the \(m=0\) term with the frequency \(\omega'_{n,0}=\gamma[ck_{n}-V K]\sim ck_{n}\gamma[1-\frac{V{\rm n}}{c}]\), which changes sign when the velocity of the lattice exceeds \(c/{\rm n}\) where the index is given by (\ref{macK}), as expected in the macroscopic limit.  The change in sign of the frequency of a wave between reference frames---as described in the introduction---can thus be understood as an indication of an underlying microscopic theory, which in our case is the presence of a lattice of point--like scattering particles.
%
%
\begin{figure}[h!]
	\includegraphics[width=14cm]{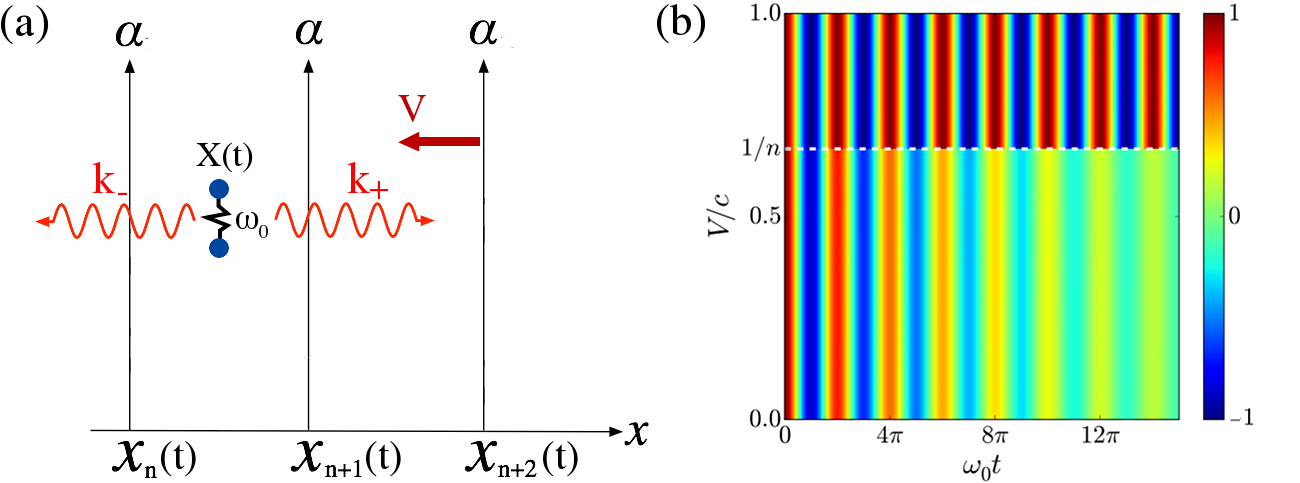}
	\caption{(a) In the frame where the lattice is in motion a small test particle is imagined to be embedded at a fixed position \(x_{0}\).  This particle has an internal degree of freedom \(X(t)\) that behaves as a simple harmonic oscillator, coupled to the field \(\varphi\).  When the lattice is at rest the coupling to the field serves to damp the internal degree of freedom, as shown in panel b.  Meanwhile when the lattice is moving, those frequencies that have changed sign between the rest frame and the moving frame serve to amplify the motion of the internal degree of freedom. (b) Plot of the motion of \(X(t)\) as a function of time and velocity computed using (\ref{damping-1}), where the lattice is treated as a medium with refractive index \({\rm n}=(1+\alpha/a)^{1/2}\) (initial conditions: \(X(0)=1\) and \(\dot{X}(0)=0\)).  Below the critical velocity \(V=c/{\rm n}\) the oscillation is damped by the coupling to the field, a damping which disappears for velocities \(V>c/{\rm n}\).  The disappearance of the damping is due to a cancellation between the energy gained/lost from the \({\rm k}_{+}\) and \({\rm k}_{-}\) modes, which both propagate left when \(V>c/{\rm n}\).  \label{moving-dipole-fig}}
\end{figure}
\par
We now explore the physical effects associated with the change of sign of the frequency between reference frames, comparing our microscopic description in terms of a lattice of scatterers and our macroscopic description in terms of a refractive index \({\rm n}\).  

\subsection{Classical dynamics of an oscillator moving through the lattice}
\par
In the macroscopic limit the medium can be characterised by the refractive index \(\rm n\) given in (\ref{macK}) and the wave equation in the frame where the medium is in motion with velocity \(-V\) takes the form
\begin{equation}
	\left[\frac{\partial^{2}}{\partial x'^{2}}-\frac{1}{c^{2}}\frac{\partial^{2}}{\partial t'^{2}}-\frac{\gamma^{2}({\rm n}^{2}-1)}{c^{2}}\left(\frac{\partial}{\partial t'}-V\frac{\partial}{\partial x'}\right)^{2}\right]\varphi(x',t')=-\kappa \delta(x'-x_{0})\dot{X}(t')\label{fieldeq}
\end{equation}
where to obtain this equation we ignored the frequency dependence of the refractive index and applied (\ref{lt}) to the rest--frame wave equation \([\partial_{x}^{2}-({\rm n}^{2}/c^{2})\partial_{t}^{2}]\varphi=0\).  The oscillating degree of freedom \(X(t)\) is coupled to the field with coupling constant \(\kappa\) and obeys the equation of a forced simple harmonic oscillator
\begin{equation}
	\ddot{X}(t')+\omega_{0}^{2}X(t')=-\kappa\dot{\varphi}(x_0,t')\label{osceqn}
\end{equation}
After performing a Fourier transform of equations (\ref{fieldeq}--\ref{osceqn}) we can immediately find the solution for \(\varphi(x,t)\) which is
\begin{equation}
	\varphi(x',t')=-\frac{\ii \kappa}{\gamma^{2}(1-\frac{{\rm n}^{2}V^{2}}{c^{2}})}\int\frac{d k}{2\pi}\int\frac{d\omega}{2\pi}\frac{\e^{\ii [k(x'-x_{0})-\omega t]}}{D(\omega,k)}\omega \tilde{X}(\omega)
\end{equation}
where \(D(\omega,k)\) is given by
\begin{equation}
	D(\omega,k)=\left(k-{\rm k}_{-}\right)\left(k-{\rm k}_{+}\right)\label{disp-mov}
\end{equation}
with
\[
	{\rm k}_{\pm}=\frac{\omega+\ii\eta}{c}\left(\frac{\pm {\rm n}-\frac{V}{c}}{1\mp\frac{{\rm n} V}{c}}\right).
\]
The two roots of \(D(\omega,k)=0\) occur where the dispersion relation for the waves in the moving frame is fulfilled.  The quantity \(\eta\) is an infinitesimal positive constant that serves to shift the zeros of \(D(\omega,k)\) into the lower half complex frequency plane.  Inserting this expression into (\ref{osceqn}) and then taking a Fourier transform gives an equation for \(\tilde{X}(\omega)\) that requires \(\omega\) to satisfy the following relation 
\begin{equation}
	\omega^{2}-\omega_{0}^{2}+\frac{\kappa^{2}\omega^{2}}{\gamma^{2}(1-\frac{{\rm n}^{2}V^{2}}{c^{2}})}\int\frac{d k}{2\pi}\frac{1}{D(\omega,k)}=0.\label{res-freq}
\end{equation}
The imaginary part of the integral in (\ref{res-freq}) determines the extent to which the coupling to the field has a damping effect on the motion of the oscillator, equivalent to a damping constant \(\Gamma=-\kappa^{2}\omega\int D(\omega,k)^{-1}dk/2\pi\).  The damping has contributions from both of the roots of (\ref{disp-mov}) and can be isolated through taking the limit \(\eta\to0\) and deforming the contour of integration to skirt around the poles on the real \(k\) axis
\begin{equation}
	\Gamma=\frac{\kappa^{2}\omega}{2\gamma^{2}(1-\frac{{\rm n}^{2}V^{2}}{c^{2}})}\sum_{\pm}\frac{\text{sign}({\rm k}_{\pm})}{\left.\frac{\partial D(\omega,k)}{\partial k}\right|_{k={\rm k}_{\pm}}}=\frac{c\kappa^{2}}{4{\rm n}}\left({\rm sign}({\rm k}_{+})-{\rm sign}({\rm k}_{-})\right)\label{damping-1}
\end{equation}
For \(|V|<c/{\rm n}\) (when the frequencies have the same sign in the moving frame and the rest frame of the medium), \({\rm k}_{+}\) and \({\rm k}_{-}\) have opposite signs and (\ref{damping-1}) is positive, meaning that the motion of the oscillator is damped by its coupling to the field.  The two terms in the brackets on the far right of (\ref{damping-1}) originate from the two solutions to the dispersion relation (\ref{disp-mov}), and evidently in this regime both solutions contribute equally to this damping.  Meanwhile, when \(|V|>c/{\rm n}\), \({\rm k}_{+}\) and \({\rm k}_{-}\) have the same sign and (\ref{damping-1}) vanishes, the coupling to the field no longer providing any damping of the oscillator.  This is because one of the two modes now amplifies the oscillatory motion with an equal magnitude to the damping due to the second mode.  The amplifying mode has a frequency of a different sign in the moving frame compared to the rest frame of the medium, due to the fact that \(|V|>c/{\rm n}\).  This behaviour of \(X(t)\) is shown in figure~\ref{moving-dipole-fig}b.
\par
\begin{figure}[h!]
	\includegraphics[width=16cm]{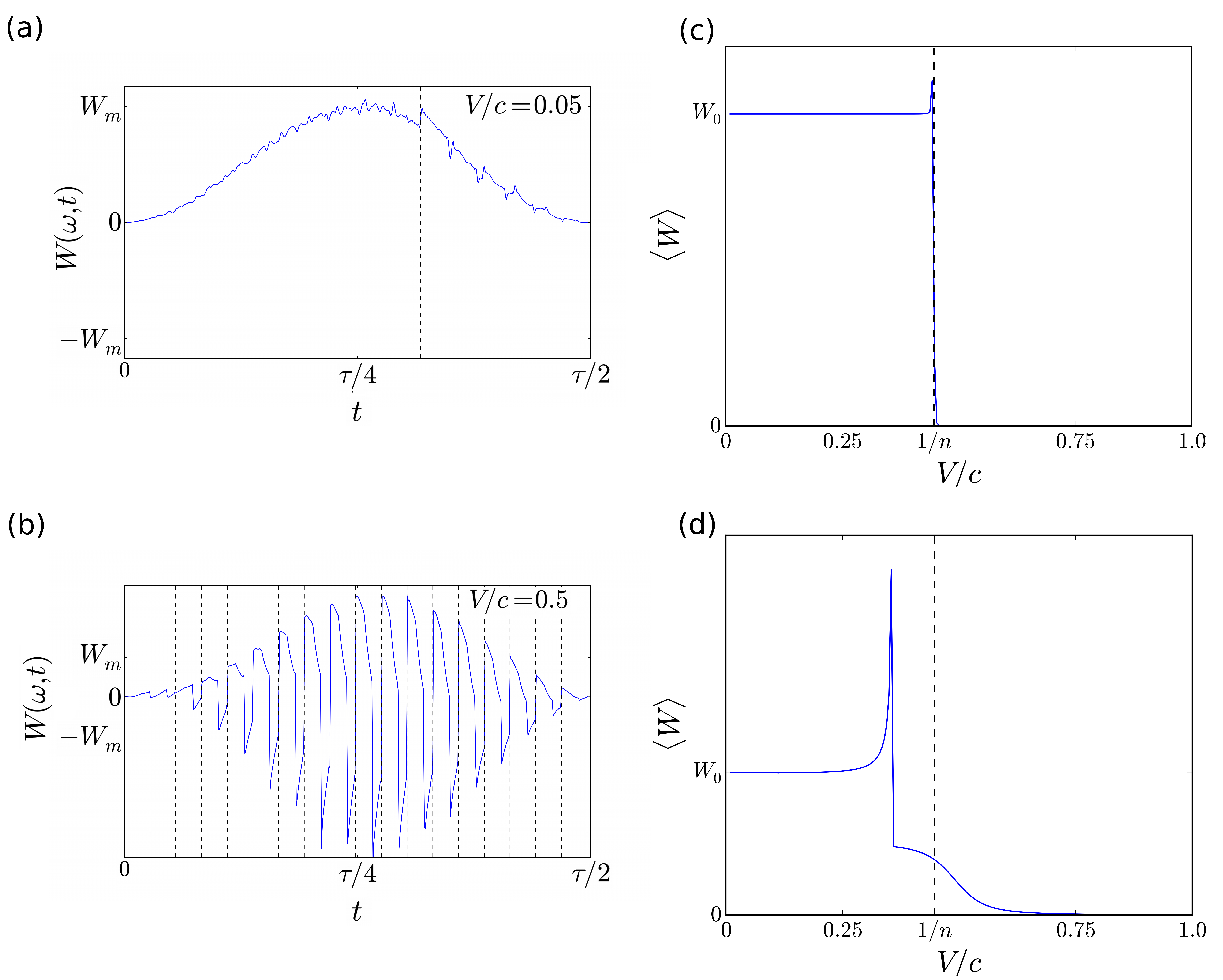}
	\caption{Work done to move an oscillator with a constant amplitude and frequency through a lattice of scatterers (\(\alpha=4.0\)).  Panels (a--b) show the work required as a function of time computed from (\ref{Weq}) for half a period of oscillation \(\tau=\pi/\omega=\pi a/c\).   The maximum value of the work required to maintain the motion of the oscillator when \(V/c=0\) is given by \(W_{m}\).  As the velocity of the oscillator is increased from zero, sharp features appear in the work as a function of time.  These are due to the collisions with the scatterers, occuring at the times shown by the vertical dashed lines.  The collisions result in the emission of sharp wavefronts that reflect off the lattice and act back on the oscillator, resulting in the sequence of sharp features in panel (a).  As the velocity is increased beyond \(V=c/{\rm n}=c/\sqrt{1+\alpha/a}=c/\sqrt{5}\) the collisions with the scatterers result in a field that acts back on the oscillator, adding and subtracting energy in equal amounts.  Panel (b) shows that in this regime the requisite work oscillates, averaging to zero as predicted in the macroscopic limit.  Panels (c--d) show two plots of the time averaged work (\ref{average_work}) as a function of velocity, with \(W_{0}\) being the average work required when the oscillator is at rest (in the macroscopic limit \(W_{0}=\Gamma\omega^{2}\)).  In panel (c) the frequency is \(\omega=0.0005 c/a\) and the work decreases in a step--like fashion as the velocity is increased above \(c/{\rm n}\) (as in the macroscopic limit shown in figure~\ref{moving-dipole-fig}), meanwhile panel (d) shows a higher oscillation frequency \(\omega=0.1c/a\) where the macroscopic limit is less applicable, and the work less rapidly reduces to zero.\label{work_figure}}
\end{figure}

When the oscillator moves through the medium faster than \(c/{\rm n}\), its radiation damping reduces to zero.  We can better understand why this is so through using our more detailed description, where we describe the medium as a lattice of scatterers.  In this case the equation of motion for the oscillator is considerably more complicated and is derived in appendix~\ref{apA}
\begin{equation}
	\ddot{X}(t')+\omega_{0}^{2}X(t')=\kappa^{2}\frac{\partial}{\partial t'}\int_{-\infty}^{\infty}dt_{0}'G(x'_{0},x_{0}',t',t_{0}')\dot{X}(t_{0}')\label{Xtlat}
\end{equation}
where the Green function \(G(x',x_{0}',t',t_{0}')\) is given by equation (\ref{ap:Gtp}), which for that particular case of \(x'=x_{0}'\) reduces to
\begin{multline}
	G(x'_{0},x_{0}',t',t_{0}')=\int_{-\infty}^{\infty}\frac{d\Omega}{2\pi}\int_{-\pi/a}^{\pi/a}\frac{dK}{2\pi}\bigg[\sum_{n}\phi_{n,K}(-\gamma V(t'-t_{0}'))\\
	-\frac{\alpha k_{0}^{2}}{a D(\Omega,K)}\sum_{n,m}\phi_{n,K}(\gamma(x_{0}'-Vt'))\phi_{m,K}(-\gamma(x_{0}'-Vt_{0}'))\bigg]\e^{-\ii\gamma(\Omega+K V)(t'-t_{0}')}\label{Gtt0}
\end{multline}
with the function \(\phi_{n,K}\) is defined by (\ref{ap:auxfun}).  The Green function (\ref{Gtt0}) depends separately on \(t'\) and \(t_{0}'\) rather than just \(t'-t_{0}'\) as in the formulae preceding (\ref{damping-1}).  This means that the damping of the oscillator is no longer simply related to the Fourier transform of (\ref{Gtt0}).  Nevertheless if we assume a fixed harmonic motion \(X(t_{0}')=\cos(\omega t_{0}')\), then the right hand side of (\ref{Xtlat}) tells us the corresponding force \(F(\omega,t')\) that would be required to maintain this motion.  The necessary work required is then given by \(W(\omega,t)=-\dot{X}(t')F(\omega,t')\)
\begin{equation}
	W(\omega,t')=-\sin(\omega t')\omega^{2}\kappa^{2}\frac{\partial}{\partial t'}\int_{-\infty}^{\infty}dt_{0}' G(x'_{0},x_{0}',t',t_{0}')\sin(\omega t_{0}').\label{Weq}
\end{equation}
If this requisite work is averaged over a time interval \(t'\in[-T/2,T/2]\) then one obtains a quantity equivalent to the damping of the oscillator as it moves through the lattice.   As \(T\to\infty\) one obtains using (\ref{Gtt0}),
\begin{align}
	\langle W(\omega)\rangle&=-\frac{\omega^{3}\kappa^{2}}{2\gamma}{\rm Im}\left[\int_{-\infty}^{\infty}\frac{dK}{2\pi}\left(\frac{1}{\left(\frac{\omega_{K}}{c}\right)^{2}-K^{2}}-\frac{\alpha}{a}\frac{(\omega_{K}/c)^{2}}{D(\omega_{K},K)\left[\left(\frac{\omega_{K}}{c}\right)^{2}-K^{2}\right]^{2}}\right)\right]\nonumber\\
	&=\frac{\alpha\omega^{3}\kappa^{2}}{4\gamma a}\sum_{n}\frac{s_{n}(\omega_{n}/c)^{2}}{\left(\frac{\partial D(\omega_{K},K)}{\partial K}\right)_{K=K_{n}}\left[\left(\frac{\omega_{n}}{c}\right)^{2}-K_{n}^{2}\right]^{2}}\label{average_work}
\end{align}
where \(s_{n}=-{\rm sign}(\partial_{\omega}D(\omega_{K},K)/\partial_{K}D(\omega_{K},K))_{K=K_{n}}\), \(\omega_{K}=\omega/\gamma-VK+\ii\eta\) and \(\omega_{n}=\omega/\gamma-VK_{n}+\ii\eta\), with \(K_{n}\) the roots of \(D(\omega_{K},K)=0\).  In figure~\ref{work_figure} the functions (\ref{Weq}) and (\ref{average_work}) are plotted to illustrate the correspondence between this lattice model of the medium, and the macroscopic description in terms of a refractive index.  As is evident from panels~\ref{work_figure}a and~\ref{work_figure}b, the fact that the damping of the oscillator is predicted to reduce to zero (\ref{damping-1}) can be attributed to the collisions with the scatterers that make up the lattice.  Each collision causes an emission of radiation which then acts back on the motion of the oscillation.  \emph{Above \(V=c/n\) this radiation acts to both damp and amplify the oscillator in equal amounts}.  In panels~\ref{work_figure}c and~\ref{work_figure}d the time averaged work is plotted for two different frequencies of oscillation and shows that for low frequencies the average work required decreases rapidly to zero for velocities \(V>c/{\rm n}\) as predicted in the limit where the medium can be described with a refractive index (\ref{macK}).  For the higher frequency the oscillator remains significantly damped above the critical velocity.
%
%
\subsection{Quantum dynamics of an oscillator moving through the lattice\label{quantum-section}}
\par
We have seen in the previous section that the damping of a classical oscillator that is being dragged through a medium reduces to zero once the velocity of dragging exceeds the phase velocity of the waves in the medium, attributed to the change in sign of one of the frequencies between reference frames.  Microscopically this can be understood in terms of the reflection of the field from the moving lattice of scatterers that make up the medium.  In this section we show that an analogous effect also occurs quantum mechanically.  When the oscillator and field are initially prepared in their ground states then the motion of the medium can lead to excitations of both the waves and the oscillator, again due to the mode with a frequency that changes sign between reference frames.  This phenomenon is similar to that presented in~\cite{horsley2012} where the oscillator was imagined to be held close to a moving surface.  However in this case we include the lattice of scatterers rather than just describe things in terms of a refractive index.
\par
To describe the system quantum mechanically we start from its Lagrangian, and to illustrate the frame independence of our results we work in the frame where the medium is at rest and the oscillator is in motion.  When the lattice of scatterers is at rest then the Helmholtz equation (\ref{eq:WaveEq}) and the equation of motion for the oscillator (\ref{osceqn}) can be derived from an action \(S=\int L dt\), where
\begin{equation}\label{eq:Lagrangian}
	L=\intop_{-\infty}^{\infty}dx\left[-\frac{1}{2}\left(\frac{\partial \varphi}{\partial x}\right)^{2}+\frac{\rho(x)}{2c^{2}}\left(\frac{\partial \varphi}{\partial t}\right)^{2}-\kappa X\delta(x-Vt)\left(\frac{\partial\varphi}{\partial t}+V\frac{\partial\varphi}{\partial x}\right)
\right]+\frac{\gamma}{2}\left[\dot{X}^{2}-\left(\frac{\omega_{0}}{\gamma}\right)^{2}X^{2}\right]
\end{equation}
where $\rho(x)=\left[1+\alpha \sum_{n}\delta(x-na)\right]$, and $X$ represents the oscillator amplitude.  The velocity dependence of (\ref{eq:Lagrangian}) was found through transforming the equation (\ref{osceqn}) into the frame where the oscillator is in motion.
\par
The Hamiltonian of the system can then be constructed from the canonical variables derived from the Lagrangian.  The canonical momentum of the field (defined in terms of the Lagrangian density \(\mathscr{L}\)) is given by
\begin{equation}\label{canmom}
	\Pi_{\varphi}=\frac{\partial\mathscr{L}}{\partial(\partial_{t}\varphi)}=\frac{\rho(x)}{c^{2}}\frac{\partial\varphi}{\partial t}-\kappa X\delta(x-Vt)
\end{equation}
and that of the oscillator by
\begin{equation}\label{canmomX}
	p_{X}=\frac{\partial L}{\partial(\partial_{t} X)}=\gamma\frac{\partial X}{\partial t}.
\end{equation}
The Hamiltonian can then be written in terms of these canonical variables via
\begin{equation}\label{Hamiltonian}
\begin{split}
	H&=\int_{-\infty}^{\infty}dx\,\Pi_{\varphi}\cdot\frac{\partial\varphi}{\partial t}+p_{X}\cdot\frac{\partial X}{\partial t}-L
	\\&=\int_{-\infty}^{\infty}dx\left\{\frac{1}{2}\left(\frac{\partial\varphi}{\partial x}\right)^{2}+\frac{c^{2}}{2\rho(x)}\left[\Pi_{\varphi}+\kappa X\delta\left(x-Vt\right)\right]^{2}+\kappa X\delta\left(x-Vt\right)V\frac{\partial\varphi}{\partial x}\right\} +\frac{1}{2\gamma}\left(p_{X}^{2}+\omega_{0}^{2}X^{2}\right)
\end{split}
\end{equation}
It is slightly difficult to make sense of this Hamiltonian because we have to---for instance---divide by \(\rho(x)\)which contains delta functions.  Where this is problematic we must understand the delta functions as the limit of a sharply peaked function.
\par
The system is quantised by rewriting the canonical variables \((\varphi,\Pi_{\varphi})\) as operators \(\hat{\varphi},\hat{\Pi}_{\varphi}\) in the Hamiltonian (\ref{Hamiltonian}) and enforcing the commutation relations
\begin{align}\label{eq:Commutation1}
	\left[\hat{\varphi}(x),\hat{\Pi}_{\varphi}(x')\right]&={\rm i}\hbar\delta(x-x')\nonumber\\
	\left[\hat{X},\hat{p}_{X}\right]&={\rm i}\hbar
\end{align}
The Hamiltonian operator (\ref{Hamiltonian}) can be split into a non--interacting part \(\hat{H}_{0}\) and an interaction term \(\hat{H}_{I}\)
\begin{equation}
	\hat{H} = \hat{H}_{0} + \hat{H}_{I} 
\end{equation}
where to leading order in \(\kappa\)
\begin{equation}
	\hat{H}_{I}=\kappa\int_{-\infty}^{\infty}dx\hat{X}\delta\left(x-Vt\right)\left[\frac{c^{2}}{\rho(x)}\hat{\Pi}_{\varphi}+V\frac{\partial\hat{\varphi}}{\partial x}\right]
\end{equation}
We work in the interaction picture~\cite{volume4}, where the time dependence of the operators is generated by \(\hat{H}_{0}\), and that of the quantum state by \(\hat{H}_{I}(t)=\exp({\rm i}\hat{H}_{0}t/\hbar)\hat{H}_{I}\exp(-{\rm i}\hat{H}_{0}t/\hbar)\).  The field operator \(\hat{\varphi}\) and its conjugate variable \(\hat{\Pi}_{\varphi}\) are written as
\begin{equation}
	\hat{\varphi}(x,t)=\sum_{n}\int_{-\pi/a}^{\pi/a}\frac{dK}{2\pi}\sqrt{\frac{\hbar c^{2}}{2\omega_{n,K}}}\left[u_{n,K}\left(x\right)e^{i\left(Kx-\omega_{n,K}t\right)}\hat{a}_{n,K}+u_{n,K}^{*}\left(x\right)e^{-i\left(Kx-\omega_{n,K}t\right)}\hat{a}_{n,K}^{\dagger}\right] \label{eq:QuantumExpansion}
\end{equation}
and
\begin{equation}
	\label{eq:MomentumExpansion}
	\hat{\Pi}_{\varphi}(x,t)=-\frac{i\rho(x)}{c^{2}}\sum_{n}\int_{-\pi/a}^{\pi/a}\frac{dK}{2\pi}\sqrt{\frac{\hbar c^{2}\omega_{n,K}}{2}}\left[u_{n,K}\left(x\right)e^{i\left(Kx-\omega_{n,K}t\right)}\hat{a}_{n,K}-u_{n,K}^{*}\left(x\right)e^{-i\left(Kx-\omega_{n,K}t\right)}\hat{a}_{n,K}^{\dagger}\right]
\end{equation}
where $\omega_{n,K}$ are the solutions to the dispersion relation of waves in the lattice (\ref{eq:BlochVector}) and the functions \(u_{n,K}\) are normalized according to (\ref{norm-cond}).  Meanwhile the oscillator operators \(\hat{X}\) and \(\hat{p}_{X}\) are written as
\begin{align}
	\hat{X}&={\rm i}\sqrt{\frac{\hbar}{2\omega_{0}}}\left(\hat{b}{\rm e}^{-{\rm i}\omega_{0}t/\gamma}-\hat{b}^{\dagger}{\rm e}^{{\rm i}\omega_{0}t/\gamma}\right)\nonumber\\
	\hat{p}_{X}&=\sqrt{\frac{\hbar\omega_{0}}{2}}\left(\hat{b}{\rm e}^{-{\rm i}\omega_{0}t/\gamma}+\hat{b}^{\dagger}{\rm e}^{{\rm i}\omega_{0}t/\gamma}\right)
\end{align}
The creation and annihilation operators obey the relations
\begin{align}
	\big[\hat{a}_{n,K},\hat{a}_{m,K'}^{\dagger}\big]&=2\pi \delta(K-K')\delta_{nm}\nonumber\\
	\big[\hat{b},\hat{b}^{\dagger}\big]&=1\label{eq:Commutation2}
\end{align}
with all other commutators equal to zero.  We expand the wave function for the system to first order in the interaction Hamiltonian, considering transitions of the system away from the ground state
\begin{equation}
	\left|\psi(t)\right\rangle=\left|0\right\rangle _{X}\otimes\left|0\right\rangle _{\varphi}+\sum_{n}\int_{-\pi/a}^{\pi/a}\frac{dK}{2\pi}\ \zeta_{n,K}\left(t\right)\left|1\right\rangle _{X}\otimes\left|1_{n,K}\right\rangle _{\varphi}
\end{equation}
where \(\zeta_{n,K}\) is the time dependent expansion coefficient which is to be determined, and \(|0\rangle_{X}\) and \(|0\rangle_{\varphi}\) are the ground states of the oscillator and the field respectively, and \(|1\rangle_{X}=\hat{b}^{\dagger}|0\rangle_{X}\) and \(|1_{n,K}\rangle_{\varphi}=\hat{a}^{\dagger}_{n,K}|0\rangle_{\varphi}\).  Assuming the interaction between the oscillator and the field is turned on for a time interval \([-T/2,T/2]\) and applying the Schr\"odinger equation and the commutators (\ref{eq:Commutation2}) the quantity $\zeta_{n,K}$ at the end of the interaction period is
\begin{align}\label{eq:rateofchange}
	\zeta_{n,K}\left(T/2\right)&=-\frac{i}{\hbar}\int_{-T/2}^{T/2}dt\,
	\left\langle 1_{n,K}\right|_{\varphi}\otimes\left\langle 1\right|_{X}\hat{H}_{I}\left(t\right)
	\left|0\right\rangle_{X}\otimes\left|0\right\rangle_{\varphi}\nonumber\\[10pt]
	&={\rm i}c\kappa\sqrt{\frac{1}{\omega_{0}\omega_{n,K}}}\bigg[\frac{\omega_{0}}{2\gamma}\int_{-T/2}^{T/2}u_{n,K}^{\star}(Vt){\rm e}^{{\rm i}(\omega_{0}/\gamma+\omega_{n,K}-VK)t}\,dt\nonumber\\
	&\hspace{5cm}-u_{n,K}^{\star}(VT/2)\sin\left((\omega_{0}/\gamma+\omega_{n,K}-VK)T/2\right)\bigg]
\end{align}
where it is assumed that the position of the oscillator at \(t=-T/2\) is an integer number of unit cells different from the position at \(t=T/2\).  If the motion begins and ends at the edge of a unit cell as defined in (\ref{field-form}) then \(T=Na/V\) where \(N\) is an odd integer. The integral over the function \(u_{n,K}(x)\) can be simplified to one over a single unit cell,
\begin{align}
	\int_{-T/2}^{T/2}u_{n,K}^{\star}(Vt){\rm e}^{{\rm i}(\omega_{0}/\gamma+\omega_{n,K}-VK)t}\,dt&=\sum_{m=-(N+1)/2}^{(N+1)/2}{\rm e}^{{\rm i}m\beta}\int_{-a/2V}^{a/2V}u_{n,K}^{\star}(Vt){\rm e}^{{\rm i}\beta Vt/a}\,dt\nonumber\\[10pt]
	&={\rm e}^{-{\rm i}\beta/2}\frac{\sin(\beta(N+1)/2)}{\sin(\beta/2)}\int_{-a/2V}^{a/2V}u_{n,K}^{\star}(Vt){\rm e}^{{\rm i}\beta Vt/a}\,dt
\end{align}
where \(\beta=[\omega_{0}/\gamma+\omega_{n,K}-VK]a/V\).

The total probability of the system having made a transition from the ground to excited state is then given by \(p_{0\to1}=\sum_{n}\int(dK/2\pi)|\zeta_{n,K}|^{2}\), and the rate at which the system makes these transitions is \(\Gamma_{0\to1}=p_{0\to1}/T\).  Taking the limit $N\rightarrow \infty$ we find that the oscillator makes transitions to the excited state at a rate
\begin{equation}
	\Gamma_{0\to1}=\frac{c^{2}\kappa^{2}\omega_{0}}{4\gamma^{2}}\sum_{n}\int_{-\pi/a}^{\pi/a}dK\frac{1}{\omega_{n,K}}\sum_{m=-\infty}^{\infty}\delta(\omega_{0}/\gamma+\omega_{n,K}-VK+2\pi mV/a)\left|\frac{1}{a}\int_{-a/2}^{a/2}u_{n,K}^{\star}(x){\rm e}^{-2\pi{\rm i} m x/a}\,dx\right|^{2}\label{emission-rate}
\end{equation}
where we applied the following representation of the periodic delta function
\[
	\lim_{N\to\infty}\left[\frac{\sin^{2}(N\beta/2)}{\sin^{2}(\beta/2)N}\right]=2\pi\sum_{m=-\infty}^{\infty}\delta(\beta+2\pi m).
\]
Notice the argument of the delta function in (\ref{emission-rate}) can only be zero for the waves that have changed the sign of their frequency between the rest frame of the medium and the rest frame of the oscillator, as discussed in section~\ref{neg-freq-sec}.  Although these `negative frequencies' are somewhat mysterious in the macroscopic limit, in this model system we have seen that they are simply a shorthand for the periodic interaction of the oscillator with the lattice of scatterers as they move past.  The integral on the right hand side of (\ref{emission-rate}) is the Fourier representation of the periodic functions \(u_{n,K}(x)\) which we find to be
\[
	\frac{1}{a}\int_{-a/2}^{a/2}u_{n,K}^{\star}(x){\rm e}^{-2\pi{\rm i} m x/a}\,dx=\frac{\alpha}{a}\frac{N^{\star}_{n,K}\sin(k_{n,K}a)k_{n,K}^{2}}{(K-2\pi m/a)^{2}-k_{n,K}^{2}}
\]
Thus the rate at which the oscillator extracts energy from the lattice is given by
\begin{equation}
	\Gamma_{0\to1}=\frac{\kappa^{2}\omega_{0}}{4\gamma^{2}}\frac{\alpha^{2}}{a^{2}}\sum_{n}\sum_{m}\left|\frac{\sin(k_{n,K_{m}}a)}{\sin(K_{m} a)}\right|\frac{k_{n,K_{m}}^{3}}{|1-V/v_{g}(\omega_{n,K_{m}})|(K^{2}_{m}-k_{n,K_{m}}^{2})^{2}}\label{emission-rate}
\end{equation}
where \(K_{m}\) is the \(m^{\rm th}\) solution to \(\omega_{0}/\gamma+\omega_{n,K}-VK=0\) (in this formula \(K\) is not restricted to the first Brillouin zone).  In the limit of small \(a\), only the first frequency band (\(n=1\)) contributes significantly to (\ref{emission-rate}), and a good approximation to the solution to the dispersion relation is \(K={\rm n}k_{1,K}\) where the index is that given in (\ref{macK}).  In this regime the rate of energy absorption reduces to
\begin{equation}
	\Gamma_{0\to1}\sim\frac{c\kappa^{2}}{4{\rm n}\gamma}\Theta\left(\frac{V{\rm n}}{c}-1\right)\label{mac-result}
\end{equation}
which is the negative part of the damping constant that we recognised in our classical treatment of this problem (\ref{damping-1}) (the factor of \(\gamma\) is present because we have worked in the frame where the oscillator is in motion).  Thus the modes that have changed the sign of their frequency between reference frames (positive in the rest frame of the medium, and negative in the oscillator's rest frame) can serve to excite a relatively moving quantum system above its ground state, even though the field is initially in its ground state.  The second mode that contributes positively to the damping in the classical result (\ref{damping-1}) does not contribute here because a quantum system cannot be reduced in energy below its ground state.   Having derived the result (\ref{mac-result}) from a model of a dielectric medium as a lattice of scatterers we can---as in the previous section---now understand these excitations of the oscillator as being due to the modified radiation damping force that the oscillator is subject to as it moves through the lattice (see figure~\ref{work_figure}).  When the velocity of the oscillator is above \(c/{\rm n}\) then the periodic interaction with the scatterers is such that the radiation reaction can equally both amplify and damp the vibration of the oscillator above the ground state (rather than just damp the motion as it does when \(V=0\)).  The amplifying part of this force then serves to excite the oscillator above the ground state.
\begin{figure}[h!]\includegraphics[height=6.5cm]{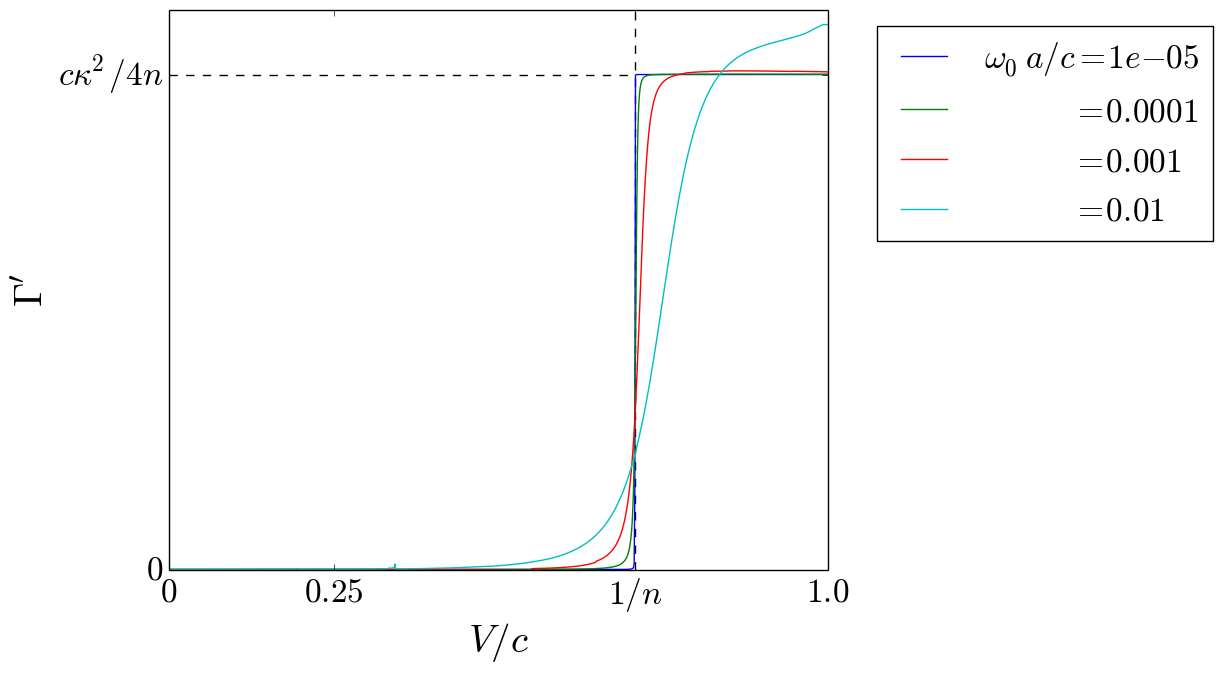}
	\caption{The absorption rate (\ref{emission-rate}) plotted for \(\alpha/a=1\) for various oscillator frequencies taken in the frame of the oscillator, $\Gamma'=\gamma\Gamma_{0\to 1}$ . For $\omega_{0}\ a/c=1.0\times 10^{-5}$, the absorption rate exhibits the step-like increase with velocity given in equation (\ref{mac-result}), representing the macroscopic limit of a homogeneous material of refractive index $n=\sqrt{1+\alpha/a}$. As $\omega_{0}\ a/c$ is increased, the effect of the lattice becomes more prominent and the velocity above which there is a non-zero absorption decreases. The effect of including the lattice is therefore to smooth of this step-like rate and allow absorption of energy for velocities much lower than the phase velocity of light in the medium originally suggested by the macroscopic limit. \label{AbsorptionRate}}
\end{figure}

\section{Summary and Conclusions}

We considered an example to illustrate the effect of frequencies that change sign between inertial reference frames: a \(1\)D scalar field coupled to a simple harmonic oscillator that moves through a medium (refractive index \({\rm n}\)) at velocity \(V\).  When the oscillator moves slower than \(c/{\rm n}\) the coupling to the scalar field has a damping effect on the oscillatory motion, while when \(V>c/{\rm n}\) we found this damping is reduced to zero.  The reason for the suppression of this damping can be understood as being due to modes that have Doppler shifted from a positive frequency in the rest frame of the medium, to a negative frequency in the rest frame of the oscillator.
\par
To better understand the physical significance of these modes we developed a simple model similar to that presented by Griffiths and Taussig~\cite{griffiths1992}, where the medium is replaced by a lattice of scatterers that in the long wavelength limit behave collectively as a refractive index \({\rm n}\).  When considered in this way, we see that the moving lattice causes a periodic forcing of the oscillator.  For velocities much below \(V=c/{\rm n}\), this periodic forcing always takes energy from the oscillatory motion.  Meanwhile for velocities above \(c/{\rm n}\) the forcing has equal contributions that are both in phase and out of phase with the oscillatory motion, corresponding to the reduced damping rate observed in the long wavelength limit.  Furthermore we found that by including the various diffracted components of the field the damping smoothly reduces as the velocity is increased, whereas in the limit where the medium is described in terms of a refractive index \({\rm n}\) the damping abruptly reduces above \(V=c/{\rm n}\).
\par
Our findings provide a more complete understanding of the `negative frequency' modes that have recently received attention in the context of quantum friction~\cite{maghrebi2013,maslovski2013} and the analogue Hawking effect~\cite{rousseaux2008,belgiorno2010}, and we can now understand them as encoding an approximation to the forcing effect of the motion of the scatterers past the observer (see figure~\ref{work_figure}).  Furthermore---to the best of the authors' knowledge---our simple model of an oscillator coupled to a wave in a moving medium is new and may be experimentally accessible.  After all, our results are not limited to the case of optics where we are restricted to small \(V/c\), but would hold for any wave theory (with some obvious modifications to the formulae, such as removing factors of \(\gamma\)).  For instance it may be interesting to investigate the phenomenon in acoustics: coupling a resonator to a moving acoustic material in which the phase velocity of sound is very low.  This could potentially be realised with an acoustic metamaterial~\cite{lu2009} constructed from a lattice of resonant elements.

\acknowledgments

The authors would like to thank T. G. Philbin and R. Churchill for useful discussions.  SARH acknowledges financial support from the EPSRC programme grant EP/1034548/1.

\appendix
\section{Emission from an oscillator moving through a lattice of scatterers\label{apA}}
\par
Suppose, as in the main text we have an oscillating degree of freedom \(X(t)\) that is moving through the lattice of scatterers described in section~\ref{sec:Microscopic-Model}.  To solve this problem we first consider the frame where the lattice is at rest.  The equation for the field of a current suddenly turned on at \(t=t_{0}\) at the point \(x=x_{0}\), (the Green function \(G(x,x_{0},t,t_{0})\)) is given by the solution of
\begin{equation}
	\left[\frac{\partial^{2}}{\partial x^{2}}-\left(1+\alpha\sum_{n}\delta(x-na)\right)\frac{1}{c^{2}}\frac{\partial^{2}}{\partial t^{2}}\right]G(x,x_{0},t,t_{0})=\delta(x-x_{0})\delta(t-t_{0})\label{ap:feq}
\end{equation}
To solve this equation we first look for the field of a source emitting at a fixed frequency \(\omega\), \(G(x,x_{0},\omega)\), which solves (\ref{ap:feq}) for the case of a monochromatic field of frequency \(\omega\), and \(\delta(x-x_{0})\) on the right hand side.  To find this quantity we define an auxiliary function \(G_{K}(x,x_{0},\omega)\)
\begin{equation}
	G(x,x_{0},\omega)=\int_{-\pi/a}^{\pi/a}\frac{dK}{2\pi}G_{K}(x,x_{0},\omega)\label{ap:GK}
\end{equation}
The quantity \(G_{K}(x,x_{0},\omega)\) is the wave produced from an array of equally spaced sources, each emitting at frequency \(\omega\) and each source differing from the previous one by a phase \(\exp(\ii K a)\),
\begin{equation}
	\left[\frac{\partial^{2}}{\partial x^{2}}+k_{0}^{2}\left(1+\alpha\sum_{n}\delta(x-na)\right)\right]
G_{K}(x,x_{0},\omega)=a\sum_{n}\e^{\ii K n a}\delta(x-x_{0}na).\label{ap:Gk}
\end{equation}
The motivation for the ansatz (\ref{ap:GK}) can be understood through considering the identity
\[
	\sum_{n}\int_{-\pi/a}^{\pi/a}\frac{dK}{2\pi}\exp(\ii K n a)\delta(x-x_{0}-na)=\frac{1}{a}\delta(x-x_{0}).
\]
From (\ref{ap:Gk}) it is clear that \(G_{K}(x+a,x_{0},\omega)=\exp(\ii K a)G_{K}(x,x_{0},\omega)\) and \(G_{K}(x,x_{0}+a,\omega)=\exp(-\ii K a)G_{K}(x,x_{0},\omega)\) which implies that it can be written in the form
\begin{equation}
	G_{K}(x,x_{0},\omega)=\sum_{n,m}g_{n,m}(K,\omega)\e^{\ii K(x-x_{0})}\e^{\frac{2\pi\ii}{a}(nx-mx_{0})}\label{ap:gkan}
\end{equation}
Substituting (\ref{ap:gkan}) in (\ref{ap:Gk}) we find
\begin{equation}
	g_{n,m}(K,\omega)
=\frac{\delta_{nm}-\frac{\alpha}{a}k_{0}^{2}\sum_{p}g_{p,m}(K,\omega)}{k_{0}^2 - \left(K+\frac{2\pi n}{a}\right)^{2}}.\label{ap:gm}
\end{equation}
which is self--consistent if
\begin{equation}
	\sum_{n}g_{n,m}(K,\omega)
=\left[1+\frac{\alpha}{a}k_{0}^{2}\sum_{p}\frac{1}{k_{0}^{2}-\left(K+\frac{2\pi p}{a}\right)^{2}}\right]^{-1}\frac{1}{k_{0}^2 - \left(K+\frac{2\pi m}{a}\right)^{2}}=\frac{1}{D(\omega,K)\left[k_{0}^{2}-\left(K+\frac{2\pi n}{a}\right)^{2}\right]}\label{ap:GK0}
\end{equation}
where \(D(\omega,K)\) is defined below.  Combining (\ref{ap:GK}--\ref{ap:GK0}) then yields the Green function at a fixed frequency
\begin{equation}
	G(x,x_{0},\omega)=\int_{-\pi/a}^{\pi/a}\frac{dK}{2\pi}\e^{\ii K(x-x_{0})}\left[\sum_{n}\phi_{n,K}(x-x_{0})-\frac{\alpha}{a}\frac{k_{0}^{2}}{D(\omega,K)}\sum_{n,m}\phi_{n,K}(x)\phi_{m,K}(-x_{0})\right]\label{ap:Gf}
\end{equation}
where
\begin{equation}
	\phi_{n,K}(x)=\frac{\e^{\frac{2\pi\ii n}{a}x}}{k_{0}^{2}-\left(K+\frac{2\pi n}{a}\right)^{2}}\label{ap:auxfun}
\end{equation}
Evidently the full Green function (\ref{ap:Gf}) breaks up into two parts.  After summation and integration the first term in the square brackets reduces to the free space Green function---it is the field of the source at \(x_{0}\) in the absence of the lattice.  The second term represents the response of the lattice to the source.  It is clear from (\ref{ap:GK0}) that the response of the lattice is strongest when \(1+\alpha k_{0}^{2}\sum_{n}[k_{0}^{2}-(K+2\pi n/a)^{2}]^{-1}=0\), which is when the lattice dispersion relation given in the main text (\ref{eq:BlochVector}) is fulfilled
\begin{equation}
	D(\omega,K)=1+\frac{\alpha}{a} k_{0}^{2}\sum_{n}\frac{1}{k_{0}^{2}-\left(K + \frac{2\pi n}{a}\right)^{2}}=1-\frac{\alpha k_{0}\sin(k_0 a)}{2\left[\cos(k_0 a)-\cos(K a)\right]}\label{ap:latdisp}
\end{equation}
\par
The time domain Green function \(G(x,x_{0},t,t_{0})\) which we set out to obtain is then given by the Fourier transform of (\ref{ap:Gf})
\begin{equation}
	G(x,x_{0},t,t_{0})=\int_{-\infty}^{\infty}\frac{d\omega}{2\pi}G(x,x_{0},\omega)\e^{-\ii\omega(t-t_{0})}\label{ap:Gft}
\end{equation}
To find the corresponding quantity in the frame where the lattice is in motion with velocity \(-V\) we perform the Lorentz transformation (\ref{lt}) of both \((x,t)\) and \((x_{0},t_{0})\).  This is
\begin{multline}
	G(x',x_{0}',t',t_{0}')=\int_{-\infty}^{\infty}\frac{d\omega}{2\pi}\int_{-\pi/a}^{\pi/a}\frac{dK}{2\pi}\e^{\ii \gamma K(x'-x_{0}'-V(t'-t_{0}'))}\bigg[\sum_{n}\phi_{n,K}(\gamma(x'-x'_{0}-V(t'-t_{0}')))\\
	-\frac{\alpha}{a}\frac{k_{0}^{2}}{D(\omega,K)}\sum_{n,m}\phi_{n,K}(\gamma(x'-Vt'))\phi_{m,K}(-\gamma(x_{0}'-Vt_{0}'))\bigg]\e^{-\ii\omega\gamma(t'-t_{0}'-V(x'-x_{0}')/c^{2})}\label{ap:Gtp}
\end{multline}
From the coupling to the oscillator given in (\ref{fieldeq}), the field in the moving lattice is thus equal to
\[
	\varphi(x',t')=-\kappa \int_{-\infty}^{\infty} dt_{0}'G(x',x_{0}',t',t_{0}')\dot{X}(t_{0'})
\]
and the motion of the oscillator obeys
\begin{equation}
	\ddot{X}(t')+\omega_{0}^{2}X(t')=\kappa^{2} \frac{\partial}{\partial t'}\int_{-\infty}^{\infty} dt_{0}'G(x'_{0},x_{0}',t',t_{0}')\dot{X}(t_{0}').\label{ap:osceqn}
\end{equation}

\end{document}